\documentclass{article}

\usepackage[nonatbib,final]{neurips_2023}

\usepackage[utf8]{inputenc} 
\usepackage[T1]{fontenc}    
\usepackage{hyperref}       
\usepackage{url}            
\usepackage{booktabs}       
\usepackage{amsfonts}       
\usepackage{nicefrac}       
\usepackage{microtype}      
\usepackage{xcolor}         
\usepackage{makecell}
\usepackage{graphicx}
\usepackage{wrapfig}
\usepackage{pifont}

\newcommand{\cmark}{\ding{51}}%
\newcommand{\xmark}{\ding{55}}%

\makeatletter
\renewcommand\section{\@startsection {section}{1}{\z@}%
                                    {-2.3ex \@plus 0.0ex \@minus -.2ex}%
                                    {1.5ex \@plus.12ex}%
                                    {\normalfont\Large\bfseries}}
\renewcommand\subsection{\@startsection{subsection}{2}{\z@}%
                                      {-1.5ex \@plus 0.0ex \@minus -.2ex}%
                                      {0.75ex \@plus 0.12ex}%
                                      {\normalfont\large\bfseries}}
\makeatother  

\title{Multimodal Data and Resource Efficient Device-directed Speech Detection with Large Foundation Models}

\author{%
  Dominik Wagner$^{1}$\thanks{Work done during an internship at Apple.}, \quad
  Alexander Churchill$^{2}$,\quad
  Siddharth Sigtia$^{2}$,\quad
  Panayiotis Georgiou$^{2}$,\quad \\
  \textbf{Matt Mirsamadi}$^{2}$\textbf{,}\quad
  \textbf{Aarshee Mishra}$^{2}$\textbf{,}\quad
  \textbf{Erik Marchi}$^{2}$\quad \\
  $^{1}$TH Nürnberg, $^{2}$Apple \\
  \texttt{dominik.wagner@th-nuernberg.de, alex.churchill@apple.com,}\\
  \texttt{sidsigtia@apple.com, panayiotis\_georgiou@apple.com,}\\
  \texttt{smirsamadi@apple.com, aarshee\_mishra@apple.com, emarchi@apple.com}
}

\begin{document}

\maketitle
\vspace{-8mm}
\begin{abstract}
\vspace{-4mm}
Interactions with virtual assistants typically start with a trigger phrase followed by a command. 
In this work, we explore the possibility of making these interactions more natural by eliminating the need for a trigger phrase. 
Our goal is to determine whether a user addressed the virtual assistant based on signals obtained from the streaming audio recorded by the device’s microphone. 
We address this task by combining 1-best hypotheses and decoder signals from an automatic speech recognition system with acoustic representations from an audio encoder as input features to a large language model (LLM). 
In particular, we are interested in data and resource efficient systems that require only a small amount of training data and can operate in scenarios with only a single frozen LLM available on a device. 
For this reason, our model is trained on 80k or less examples of multimodal data using a combination of low-rank adaptation and prefix tuning. 
We compare the proposed system to unimodal baselines and show that the multimodal approach achieves lower equal-error-rates (EERs), while using only a fraction of the training data. 
We also show that low-dimensional specialized audio representations lead to lower EERs than high-dimensional general audio representations.  
\end{abstract}
\vspace{-6mm}
\section{Introduction}
\vspace{-3mm}
Speech-based virtual assistants allow users to interact with devices such as  phones, watches, and loudspeakers via voice commands. 
To distinguish audio that is directed towards a device from background speech, a trigger phrase or the press of a button usually precedes the user command \cite{voicetrigger23}. 
The problem of detecting a trigger phrase is referred to as, wake-word detection \cite{jose20_interspeech,ghosh22_interspeech}, voice trigger detection \cite{sigtia18_interspeech,sigtia20vtd}, or keyword spotting \cite{sainath15kws,michaely17kws,cornell22ddsd,ng23kws}. 
To create a more natural conversation flow, subsequent commands after the initial interaction should not require the trigger phrase. 
Device-directed speech detection is concerned with determining whether a virtual assistant was addressed or not, without a trigger cue preceding the voice command at all times \cite{shriberg12_interspeech,mallidi18_interspeech,garg22_interspeech}.
Device-directed speech detection systems are exposed to information from all kinds of in-domain (voice commands) and out-of-domain (e.g. background speech, ambient sounds, appliances etc.) signals. 
Previous works use a combination of acoustic and lexical features to encode the relevant information in those signals \cite{shriberg12_interspeech,mallidi18_interspeech,gillespie20dd,sato23multimodal,bekal22_interspeech}. 

Recent studies have extended LLMs with the ability to process non-lexical input modalities, such as audio and video data \cite{mokady2021clipcap,driess2023palme,fathullah2023prompting,gong2023ltu,kim2023prefix,deshmukh2023pengi}. 
Inspired by these efforts, we explore a LLM-based multimodal model to differentiate between directed and non-directed audio in interactions with a virtual assistant. 
Our goal is to determine whether the user addressed the assistant using signals obtained from the streaming audio captured by the device's microphone. 

The proposed model uses acoustic features obtained from a pretrained audio encoder in combination with decoder signals, such as acoustic cost, as well as 1-best hypotheses from an ASR system. 
The acoustic features and decoder signals are represented as learnable fixed-length prefixes, which are concatenated with the token embeddings of the 1-best hypotheses (cf. Figure \ref{fig:arch}). 
The system is optimized to generate decisions about device-directedness by jointly learning from all modalities using a combination of prefix tuning \cite{liang21prefixtuning} and low-rank adaptation (LoRA) \cite{hu2022lora}.

We analyze this task in a scenario in which (1) only a limited amount of training data is available and (2) only a pretrained LLM with frozen weights is usable on a resource-constrained device (e.g. a smartphone). 
Furthermore, we compare the effectiveness of high-dimensional representations obtained from a large generic audio foundation model with lower-dimensional representations from a small audio encoder trained on in-domain data. 

\vspace{-4mm}
\section{Feature Extraction}\label{sec:feat}
\vspace{-3mm}
\paragraph{1-best Hypotheses and ASR Decoder Signals}
The text part of the data was transcribed with an on-device joint CTC-attention based end-to-end speech recognition system \cite{kim17ctcatt} trained on in-domain data, comparable to the one used in \cite{bleeker23_interspeech}. 
Inspired by \cite{shriberg12_interspeech,mallidi18_interspeech}, we extract 4 additional utterance-level signals that are generated by a decoder based on weighted finite-state transducers \cite{miao15eesen}. 
For the most likely hypothesis in the N-best set of hypotheses, we extract the average of the graph cost associated with each word in the hypothesis, the average of the acoustic cost, and the average of the word-level posterior confidence scores. 
The graph cost is the sum of language model cost, transition probabilities, and pronunciation cost \cite{povey12wfst}.
The acoustic cost is the negative log-likelihood of the tokens from the decoder. 
Additionally, we include the average number of alternative word options for each word in the 1-best hypothesis. 
Finally, we scale the feature values along each signal dimension into the unit interval $\left[ 0, 1\right]$ across the dataset. 
\vspace{-4mm}
\paragraph{Audio Representations}
We compare two pretrained models as backbones to extract audio representations. 
The first model is the medium version of Whisper (769M parameters) \cite{radford2022robust}.  
Whisper is expected to generalize well across domains and languages, since it was trained on 680k hours of speech data and is therefore well-suited for our task. 
We extract 1024-dimensional representations at the last encoder layer of Whisper. 
Additionally, we explore a specialized and lightweight on-device model for acoustic feature extraction. 
We choose the Unified Acoustic Detector (UAD) for false trigger mitigation described in \cite{rudovic23sdsd} as an alternative feature extractor. 
The model is trained to detect unintended invocations of devices such as smartphones. 
It has $\approx$6 million parameters and consists of a shared transformer-based encoder \cite{vaswani17attention}, followed by task-specific classification heads. 
We extract 256-dimensional representations at one of the task-specific classification heads. 
\vspace{-4mm}
\section{Method}
\vspace{-3mm}
\begin{wrapfigure}{r}{0.65\textwidth}
  \begin{center}
    \vspace*{-8mm} 
    \hspace*{-5mm}  
        \includegraphics[width=0.7\textwidth]{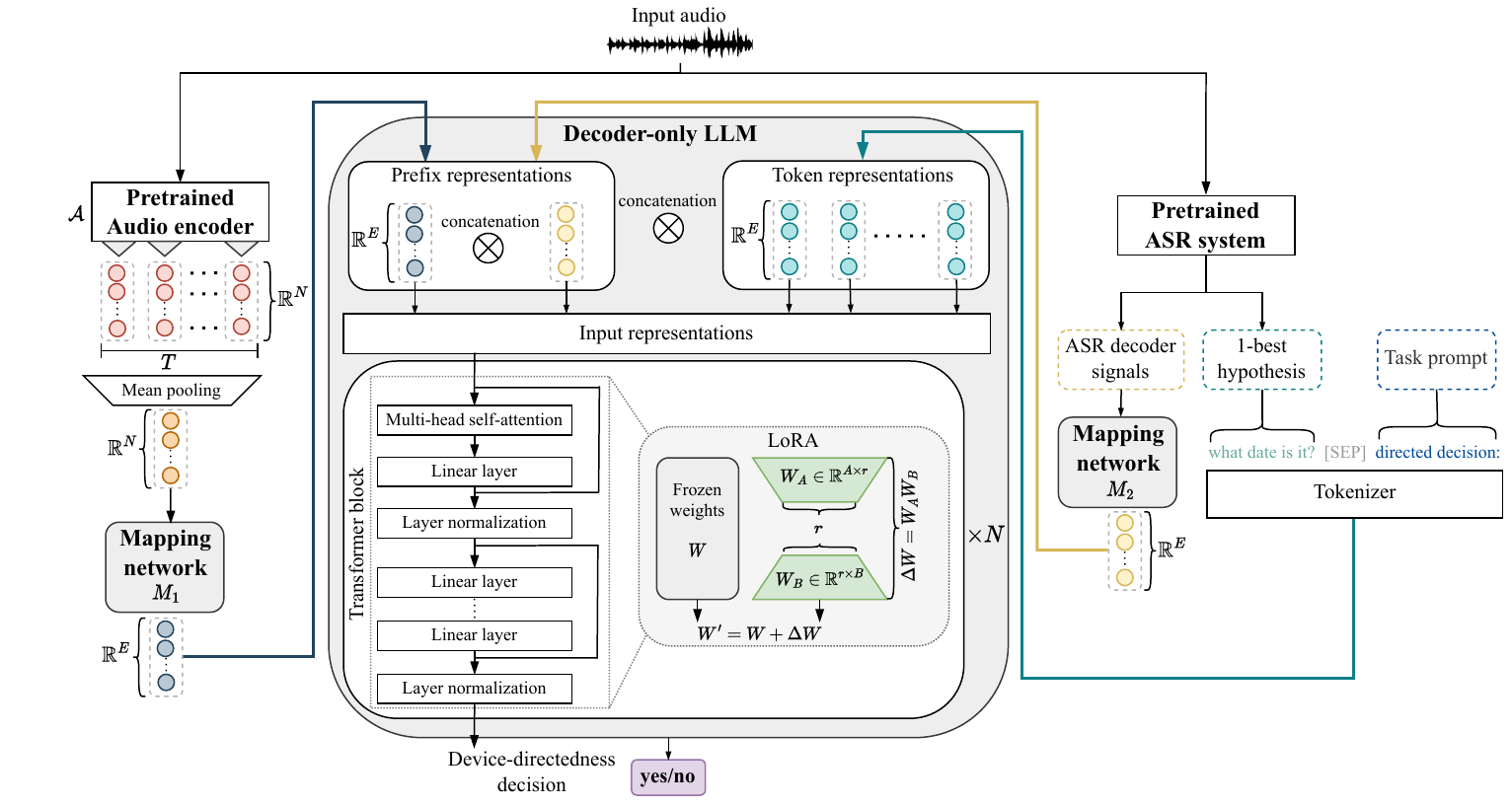}
  \end{center}
    \vspace{-5mm}
    \caption{Architecture of the multimodal system. The weights of the LoRA modules are trained along with the weights of $M_1$ and $M_2$. All other components remain frozen. }
    \label{fig:arch}
    \vspace{-4.5mm}
\end{wrapfigure}
Our system consists of three main components (cf. Figure \ref{fig:arch}). 
The first main component is a frozen audio encoder (either Whisper or UAD), which extracts a sequence of latent representations in $\mathbb{R}^{N}$ from the input audio. 
The second main component comprises two feedforward mapping networks, $M_1$ and $M_2$, which translate the extracted audio features and the utterance-level ASR decoder signals into the latent space of the token embeddings. 
The third main component is a decoder-only LLM that generates text given the prefix representations obtained via $M_1$ and $M_2$ as well as the 1-best ASR hypotheses, and a text prompt (i.e., ``directed decision:'' in Figure \ref{fig:arch}).

The task is to generate a decision on whether an unseen utterance is directed towards a device or not. 
The model is trained on multimodal data that contains $L$ examples of audio, ASR decoder signals, and 1-best hypotheses $\left\{(x^i, d^i, t^i)\right\}_{i=1}^L$. The 1-best hypotheses are represented by a sequence of tokens $t^i =(t^i_1, \ldots, t^i_l)$, which are padded to a maximum length $l$. 
The input waveform $x^i$ is transformed into log-magnitude Mel spectrogram features via the transformation $\mathcal{F}$. 
The spectrogram feature input $\mathcal{F}(x^i)$ is then processed to obtain a sequence of embeddings in $\mathbb{R}^{N}$ of length $T$ using the audio encoder $\mathcal{A}$ (either Whisper or UAD). 
Mean pooling is applied to these representations along the time dimension to generate a single vector in $\mathbb{R}^{N}$ per utterance. 
$M_1$ is then used to map the aggregated embedding to a vector in the prefix token space of the LLM:  
\begin{equation}
    a^i = M_1 \left(\frac{1}{T} \sum_{t=1} ^T h_t \right) \in \mathbb{R}^{1 \times E}, \; \mathcal{A} \left(\mathcal{F} (x^i ) \right)=[h_1 \cdots h_T]^\top \in \mathbb{R}^{T \times N}. 
\end{equation}
The resulting representation $a^i$ has the same dimensionality $E$ as the token embeddings. 
$M_2$ is used to generate a latent prefix $b^i$ in $\mathbb{R}^E$ for the decoder signal features $d^i$.  
The audio prefix $a^i$ and the decoder signal prefix $b^i$ are then concatenated to the token embeddings of the corresponding 1-best hypothesis $t^i$, and the concatenated input features are presented to the LLM. 
The training objective is to predict directedness tokens conditioned on the prefix, the decoder signals, and the 1-best hypothesis tokens in an autoregressive fashion. 
We utilize cross entropy loss to train the parameters $\theta$ of the model:
\vspace{-2mm}
\begin{equation}
    \mathcal{L}_{\theta}=-\sum_{i=1}^L \sum_{j=1}^l \log p_\theta\left(t_j^i \mid a^i, b^i, t_1^i, \ldots, t_{j-1}^i\right).
\end{equation}
During inference, the device-directedness decision is made based on the score $p_\theta\left(Y=yes \mid c \right)$, where $Y$ is a discrete random variable that can take one of $m$ tokens $y_1, ... , y_m$ from the vocabulary $\mathcal{V}$, and $p_\theta\left(Y=yes \mid c \right) + p_\theta\left(Y=no \mid c \right) \simeq 1$. 
The context $c$ is determined by the multimodal features, i.e., $c =(a^i, b^i, t_1^i, \ldots, t_{j-1}^i)$. 
\vspace{-4mm}
\paragraph{Large Language Models}
We focus on decoder-only LLMs, since this architecture choice has demonstrated stronger capabilities \cite{brown2020gpt3} than encoder-only and encoder-decoder systems, such as BERT \cite{devlin-etal-2019-bert} and T5 \cite{raffel20t5}, on a wide range of tasks. 
We compare the 7B parameter versions of Falcon \cite{falcon2023} and RedPajama \cite{together2023redpajama} in our experiments. 
\vspace{-4mm}
\paragraph{Mapping Networks}
The mapping networks $M_1$ and $M_2$ translate between the latent space of the audio encoder and the lexical embedding space of the LLM. 
All audio features and ASR decoder signals are transformed into $\mathbb{R}^{E}$ sized prefixes, where $E$ is the latent dimension of the LLM. 
Both mapping networks share the same architecture, consisting of one hidden linear layer with \nicefrac{E}{2} units and hyperbolic tangent activation.  The models are trained with a dropout \cite{srivastava14dropout} probability of 10\%. 
\vspace{-4mm}
\paragraph{Low-rank Adaptation}
We employ low-rank adaption (LoRA) \cite{hu2022lora} to finetune the LLM without directly changing its weights. 
In the LoRA method, weights of dense layers in large pretrained models are summed with linear low-rank adapter modules. 
These adapter modules are small trainable matrices, which are included into the architecture and optimized on behalf of the underlying LLM weights. 
We attach adapter modules to the query $q$ and value $v$ projection matrices, as well as the dense layers $d$ of each transformer block. 
We employ the configuration $r_q = r_v = d = 8$, $\alpha = 32$ and train the adapters with a dropout probability of 10\%. 
The parameter $r$ is the rank of the adaptation matrices, and $\alpha$ is a scaling factor to adjust the magnitude of the adaptation. 
The LoRA approach allows us to use less training data \cite{kaplan2020scaling} and enables the reuse of a generic LLM deployed on a device. 
\vspace{-4mm}
\paragraph{Unimodal Baselines}
Unimodal versions of our framework are trained by providing text, decoder signals, or audio representations as the only input source to the LLM. 
In the \textit{text-only} variant, the mapping networks $M_1$ and $M_2$ are removed, and the only input features are the 1-best hypotheses of the ASR system (cf. Figure \ref{fig:arch}). 
In the \textit{audio-only} variant, the decoder signals including $M_2$ and the 1-best hypotheses are removed from the system. 
The \textit{decoder-signal-only} system relies only on the decoder signal input, which is transformed via $M_2$. 
Hence, $M_1$ and the 1-best hypotheses are removed from the overall system. 
\vspace{-4mm}
\section{Experiments}
\vspace{-3mm}
\begin{table}[t]
\setlength{\tabcolsep}{6.5pt}
    \caption{\footnotesize{Comparison EERs on the evaluation set. ``Uni'' refers to unimodal experiments, and ``Multi'' refers to multimodal experiments. ``Modality'' indicates the modalities used in the experiment ($t$ = text, $a$ = audio, $b$ = decoder signals). ``Train Size'' shows the number of training examples used in the experiment. ``\# Param'' is the number of trainable parameters. We report the sum of the parameters of the mapping networks and LoRA.}}
    \vspace{-2mm}
  \centering \def\arraystretch{0.96} \small
  \scalebox{0.87}{%
\begin{tabular}{ cccc|ccc|ccc } 
 \hline
                 &                &                 &                         &  \multicolumn{3}{c|}{\textbf{Falcon 7B}} &  \multicolumn{3}{c}{\textbf{RedPajama 7B}}\\ 
  \textbf{Experiment} & \textbf{LoRA} & \textbf{Modality}  & \makecell{ \textbf{Train} \\ \textbf{Size}} & \makecell{ \textbf{\#} \\ \textbf{Param} }& \makecell{ \textbf{EER} \\ \textbf{Whisper} } & \makecell{ \textbf{EER} \\ \textbf{UAD} } & \makecell{ \textbf{\#} \\ \textbf{Param} } & \makecell{ \textbf{EER} \\ \textbf{Whisper} } & \makecell{ \textbf{EER} \\ \textbf{UAD} }\\
  \hline
  Uni 1     &   \cmark       & $t$                  & 80k  & 16M & 12.97\% & 12.97\% & 17M & 12.90\% &  12.90\% \\
  Uni 2     &    \cmark      & $a$                  & 80k  & 29M & 10.45\% & 9.31\%  & 27M & 10.78\% & 8.99\% \\
  Uni 3     &    \cmark      & $b$                  & 80k  & 26M & 36.90\% & 36.90\% & 25M & 35.04\% & 35.04\% \\
  \hline
  Multi 1    &   \cmark        & $t$, $b$         & 80k  & 26M & 13.39\% & 13.39\% & 25M    & 12.86\% & 12.96\% \\
  Multi 2    &    \cmark        & $a$, $b$         & 80k  & 39M & 14.94\% & 9.92\%   & 35M   & 14.80\% & 10.71\% \\
  Multi 3    &    \cmark        & $t$, $a$         & 80k  & 29M & 9.96\% & 8.76\%   & 27M    & 9.89\% & 8.44\% \\
  Multi 4    &   \cmark         & $t$, $a$, $b$    & 80k  & 39M & 8.80\% & \textbf{8.23\%} & 35M & 9.45\% & 8.52\% \\
  \hline
  Multi 5 \scriptsize{(frozen LLM)}    &  \xmark     & $t$, $a$, $b$    & 80k  & 23M & 10.52\% & 11.49\% & 18M & 10.90\% & 12.26\% \\
  \hline
  Multi 4.1   &   \cmark          & $t$, $a$, $b$ & 40k  & 39M & 10.19\% & 8.38\%    & 35M & 10.20\% & 8.47\% \\
  Multi 4.2   &   \cmark          & $t$, $a$, $b$ & 20k  & 39M & 10.67\% & 9.05\%    & 35M & 10.91\% & 8.84\% \\
  Multi 4.3   &   \cmark          & $t$, $a$, $b$ & 10k  & 39M & 11.71\% & 8.84\%    & 35M & 11.66\% & 9.69\% \\
  Multi 4.4   &   \cmark          & $t$, $a$, $b$ & 5k   & 39M & 12.76\%  & 9.77\%   & 35M & 12.11\% & 9.65\% \\
  Multi 4.5   &   \cmark          & $t$, $a$, $b$ & 1k   & 39M & 15.39\%  & 12.56\%   & 35M &17.09\% & 11.87\% \\
  \hline
\end{tabular}
}
  \label{tab:exp}
  \vspace{-7mm}
\end{table}
\paragraph{Data}
The full training data is a balanced set of $\approx$40k directed utterances and $\approx$40k non-directed utterances, similar to the set used in \cite{rudovic23sdsd} and \cite{dighe23a2i}. 
The evaluation data is a combined set of two in-house corpora with $\approx$14k device-directed utterances and $\approx$23k non-directed utterances. 
The total duration of the evaluation data is $\approx$35 hours. 
Approximately 29\% of the device-directed training examples start with a trigger phrase and $\approx$12\% of the device-directed evaluation utterances start with a trigger phrase. 
The remaining device-directed utterances are triggerless interactions with a virtual assistant. 
All utterances in the training and evaluation data are randomized and anonymized. 
The dataset statistics are summarized in Table \ref{tab:data} of Appendix \ref{app:data}. 
\vspace{-4mm}
\paragraph{Results and Discussion}
The equal-error-rates (EERs) for our experiments are summarized in Table \ref{tab:exp}. 
The unimodal baselines  (Uni 1-3) are the \textit{text-only} ($t$), \textit{audio-only} ($a$), and \textit{decoder-signal-only} ($b$) versions of the proposed system. 
Using only the audio modality (Uni 2) yields lower EERs than using only the text modality (Uni 1), irrespective of the underlying LLM and audio encoder. 
Furthermore, using the specialized UAD representations leads to lower EERs than using Whisper representations in experiment Uni 2. 
Decoder signals (Uni 3) provide the weakest overall signal ($EER=36.90\%$ with Falcon and $EER=35.04\%$ with RedPajama). 
The best system configuration (Multi 4) uses all 80k available training examples and combines information from text, audio, as well as decoder signals. 
Multi 4 with Falcon shows an EER of 8.80\% using Whisper as the audio encoder and an EER of 8.23\% with the UAD backbone, which translates to relative improvements of $\approx$16\% and $\approx$12\% over the corresponding \textit{audio-only} models (Uni 2). 
In Multi 5, only $M_1$ and $M_2$ are trained (i.e., the underlying LLM is frozen and no LoRA modules are attached). 
This configuration shows worse results than Multi 4, indicating that training the mapping networks alone is not sufficient to achieve low EERs. 
The experiments Multi 4.1 to Multi 4.5 are the same as Multi 4 but with a stepwise reduction of the training data (from 40k examples to 1k examples). 
The multimodal system with Falcon and the UAD backbone trained on only 10k examples (Multi 4.3) still performs better than the \textit{audio-only} model trained on 80k examples (EERs of 8.84\% and 9.31\%). 
This is not the case when Whisper representations are used instead (EERs of 11.71\% and 10.45\%). 
Additional experiments showing the impact of using the smallest (39M) and largest (1.5B) versions of Whisper as audio feature extractors can be found in Table \ref{tab:abl_whsp} of Appendix \ref{app:add_exp}. 

In contrast to other systems for device-directedness detection \cite{mallidi18_interspeech,gillespie20dd,rudovic22sdsd}, our approach requires only a small amount of training data. 
We see that audio representations from a pretrained Whisper model perform well in this low data environment. 
However, the model can be further improved by replacing these unspecialized representations with specialized ones obtained from the smaller UAD model. 
This effect is amplified in very low data environments (see Multi 4.1-4.5). 
While we observe a strong EER increase with less training data (e.g. from 8.80\% with 80k examples to 15.39\% with 1k examples using Falcon) when Whisper representations are used, the EER increase is less pronounced with UAD representations (e.g. from 8.23\% with 80k examples to 12.56\% with 1k examples using Falcon). 
We hypothesize that in low data environments the model relies more on what it already knows (i.e., the acoustic information encoded in the in-domain UAD model) and the amount of training data is not sufficient to learn how the acoustic information encoded in unspecialized representations can be utilized accordingly. 
\vspace{-5mm}
\section{Conclusions}
\vspace{-3.7mm}
In this work, we described a multimodal model to distinguish device-directed utterances from background speech. 
Our approach made use of knowledge encoded in pretrained foundation models and effectively combined decoder signals with audio and lexical information. 
The system can be trained on small amounts of data and operates in scenarios, where only a single frozen LLM is available on a resource-constrained device. 
We achieved lower EERs than unimodal baselines, while using only a fraction of the training data. 
Furthermore, low-dimensional audio representations from a small specialized feature encoder outperformed high-dimensional general representations from a larger audio foundation model and showed more stable results in environments with very low data availability (i.e., $<$80k utterances). 


\bibliographystyle{ieeetr}
{
\small
\bibliography{refs}
}
\newpage
\appendix
\section*{Appendix}
\section{Acknowledgements}
We would like to express our sincere gratitude to John Bridle, Pranay Dighe, Sachin Kajarekar, Oggi Rudovic, Ahmed Tewfik and Barry Theobald for their support and their comprehensive feedback on this work. 
We also thank Seanie Lee for the numerous helpful discussions. 

\section{Datasets}\label{app:data}
\begin{table}[h]
\setlength{\tabcolsep}{3pt}
  \caption{Summary of the data used in our experiments.}
  \label{tab:data}
  \centering
  \begin{tabular}{ccccccccc}
    \toprule
    \textbf{Label}     & \multicolumn{2}{c}{\makecell{\textbf{Examples} \\ (\#)}} & \multicolumn{2}{c}{\makecell{\textbf{Total Duration} \\ (hours)}} & \multicolumn{2}{c}{\makecell{\textbf{Duration per Utterance} \\ (seconds)}}  & \multicolumn{2}{c}{\makecell{\textbf{Words per Utterance} \\ (\#)}} \\
              &  Train & Eval &  Train & Eval  &  Train & Eval &  Train & Eval \\          
    \midrule
    Directed & 40568 & 14396 & 58.88 & 12.05     & 5.22$\pm$6.97 & 3.01$\pm$1.89     & 5.34$\pm$3.26 & 5.50$\pm$3.64 \\
    Non-directed & 40062  & 22958 & 67.31 & 23.37 & 6.04$\pm$5.33 & 3.66$\pm$3.67 & 6.48$\pm$9.29 & 9.42$\pm$17.28   \\
    \midrule
    Combined & 80630  & 37354 & 126.19 & 35.42 & 5.63$\pm$6.22 & 3.41$\pm$3.12 & 5.90$\pm$6.97 & 7.89$\pm$13.81 \\
    \bottomrule
  \end{tabular}
\end{table}

\section{Examples}\label{app:ex}
% ...

\begin{table}[h]
    \setlength{\tabcolsep}{0pt}
  \caption{Examples of 1-best hypotheses of device-directed and non-directed utterances.}
  \label{tab:examples}
  \centering
  \begin{tabular*}{\linewidth}{@{\extracolsep{\fill}} ll}
    \toprule
    Directed     & ``Set an alarm for 8 AM''      \\
    Directed     & ``Tell me a joke''      \\
    Directed     & ``What's the temperature''      \\
    \midrule
    Non-directed & ``Excellent thank you very much''       \\
    Non-directed & ``Can we talk''      \\
    Non-directed & ``I was trying to do it''       \\
    \bottomrule
  \end{tabular*}
\end{table}

\section{Additional Experiments}\label{app:add_exp}
\begin{table}[h]
\setlength{\tabcolsep}{5pt}
    \caption{Impact of changing the size of the Whisper audio foundation model. ``Whisper Tiny'' has $\approx$39M parameters and the audio representation dimension is $\mathbb{R}^{384}$. ``Whisper Large'' has $\approx$1.5B parameters and the audio representation dimension is $\mathbb{R}^{1280}$. The LoRA configuration is the same as in Table \ref{tab:exp} ($r_q = r_v = d = 8$, $\alpha = 32$).}
  \centering \def\arraystretch{1} \small
\begin{tabular}{ ccc|cc|cc } 
 \hline
                 &                                 &                         &  \multicolumn{2}{c|}{\textbf{Falcon 7B}} &  \multicolumn{2}{c}{\textbf{RedPajama 7B}} \\ 
  \textbf{Experiment} & \textbf{Modality}  & \makecell{ \textbf{Train} \\ \textbf{Size}} & \makecell{ \textbf{EER} \\ \textbf{Whisper Tiny} } & \makecell{ \textbf{EER} \\ \textbf{Whisper Large} } & \makecell{ \textbf{EER} \\ \textbf{Whisper Tiny} } & \makecell{ \textbf{EER} \\ \textbf{Whisper Large} } \\
  \hline
  Uni 2                 & $a$           & 80k   & 14.66\%     & 10.14\% & 13.84\% & 9.76\%  \\
  \hline
  Multi 4               & $t$, $a$, $b$ & 80k   & 10.56\%     & 10.03\% & 9.75\% & 9.02\%  \\
  \hline
\end{tabular}
  \label{tab:abl_whsp}
\end{table}

\begin{table}[t]
\setlength{\tabcolsep}{7pt}
    \caption{Alternative LoRA configuration for Falcon 7B and RedPajama 7B. LoRA modules are only attached to $q$ and $v$  and $r_q=r_v=64, \alpha=16$ is used. }
  \centering \def\arraystretch{1} \small
\begin{tabular}{ ccc|ccc|ccc } 
 \hline
                 &                                 &                         &  \multicolumn{3}{c|}{\textbf{Falcon 7B}} &  \multicolumn{3}{c}{\textbf{RedPajama 7B}}\\ 
  \textbf{Experiment} & \textbf{Modality}  & \makecell{ \textbf{Train} \\ \textbf{Size}} & \makecell{ \textbf{\#} \\ \textbf{Param} } & \makecell{ \textbf{EER} \\ \textbf{Whisper} } & \makecell{ \textbf{EER} \\ \textbf{UAD} } & \makecell{ \textbf{\#} \\ \textbf{Param} }& \makecell{ \textbf{EER} \\ \textbf{Whisper} } & \makecell{ \textbf{EER} \\ \textbf{UAD} }\\
  \hline
  Uni 1               & $t$                  & 80k  & 19M  & 12.59\% & 12.59\% & 33M & 12.88\% & 12.88\%  \\
  Uni 2               & $a$                  & 80k  & 32M  & 10.53\% & 12.52\% & 43M & 10.92\% & 9.16\%   \\
  \hline
  Multi 3               & $t$, $a$         & 80k  & 32M  & 9.33\% & 9.36\%   & 43M & 9.44\% & 8.45\%    \\
  Multi 4               & $t$, $a$, $b$ & 80k  & 42M     & 10.13\%     & 10.55\% & 51M & 9.72\%  & 9.75\% \\
  \hline
  Multi 4.1               & $t$, $a$, $b$ & 40k  & 42M  & 9.97\% & 11.10\%     & 51M & 9.86\% & 9.19\%  \\
  Multi 4.2               & $t$, $a$, $b$ & 20k  & 42M  & 11.07\% & 11.55\%    & 51M & 10.63\% & 9.73\%  \\
  Multi 4.3               & $t$, $a$, $b$ & 10k  & 42M & 10.91\% & 12.96\%     & 51M & 12.00\% & 9.95\%  \\
  Multi 4.3               & $t$, $a$, $b$ & 5k  & 42M & 12.53\% & 12.49\%        & 51M & 12.08\% & 10.81\%  \\
  Multi 4.5               & $t$, $a$, $b$ & 1k  & 42M  & 17.63\% & 13.36\%     & 51M & 14.71\% & 14.17\%  \\
  \hline
\end{tabular}
  \label{tab:abl}
\end{table}

\begin{table}[t]
\setlength{\tabcolsep}{2pt}
    \caption{Alternative LoRA configurations for Falcon 7B. Note that the configuration in the middle column ($r_q=r_v=d=8, \alpha=32$) is the same as in Table \ref{tab:exp}. }
  \centering \def\arraystretch{1} \small
\begin{tabular}{ ccc|cc|cc|cc } 
 \hline
                 &                                 &                         &  \multicolumn{2}{c|}{$r_q=r_v=8, \alpha=32$} &  \multicolumn{2}{c|}{$r_q=r_v=d=8, \alpha=32$} &  \multicolumn{2}{c}{ $r_q=r_v=d=64, \alpha=16$}\\ 
  \textbf{Experiment} & \textbf{Modality}  & \makecell{ \textbf{Train} \\ \textbf{Size}}  & \makecell{ \textbf{EER} \\ \textbf{Whisper} } & \makecell{ \textbf{EER} \\ \textbf{UAD} } & \makecell{ \textbf{EER} \\ \textbf{Whisper} } & \makecell{ \textbf{EER} \\ \textbf{UAD} } & \makecell{ \textbf{EER} \\ \textbf{Whisper} } & \makecell{ \textbf{EER} \\ \textbf{UAD} }\\
  \hline
  Uni 1               & $t$                  & 80k  & 12.53\% & 12.53\%  & 12.97\% & 12.97\%  & 12.47\% &  12.47\% \\
  Uni 2               & $a$                  & 80k  & 10.33\% & 13.81\%  & 10.45\% & 9.31\%   & 10.95\% & 9.25\% \\
  Uni 3               & $b$                  & 80k  & 34.24\% & 34.24\% & 36.90\% & 36.90\%  & 35.42\% & 35.42\% \\
  \hline
  Multi 1               & $t$, $b$         & 80k  & 13.19\% & 13.19\%    & 13.39\% & 13.39\%  & 12.99\% & 12.99\% \\
  Multi 2               & $a$, $b$         & 80k & 16.85\% & 10.49\%     & 14.94\% & 9.92\%  & 13.48\% & 10.35\% \\
  Multi 3               & $t$, $a$         & 80k & 9.09\% & 9.08\%       & 9.96\% & 8.76\%    & 9.51\% & 8.00\% \\
  Multi 4               & $t$, $a$, $b$    & 80k   & 9.17\% & 10.92\%    & 8.80\% & 8.23\%   & 9.09\% & 8.00\% \\
  \hline
\end{tabular}
  \label{tab:test}
\end{table}

\textbf{\begin{table}[t]
\setlength{\tabcolsep}{2pt}
    \caption{Alternative LoRA configuration for RedPajama 7B. Note that the configuration in the middle column ($r_q=r_v=d=8, \alpha=32$) is the same as in Table \ref{tab:exp}.}
  \centering \def\arraystretch{1} \small
\begin{tabular}{ ccc|cc|cc|cc } 
 \hline
                 &                                 &                         &  \multicolumn{2}{c|}{$r_q=r_v=8, \alpha=32$} &  \multicolumn{2}{c|}{$r_q=r_v=d=8, \alpha=32$} &  \multicolumn{2}{c}{$r_q=r_v=d=64, \alpha=16$}\\ 
  \textbf{Experiment} & \textbf{Modality}  & \makecell{ \textbf{Train} \\ \textbf{Size}} &  \makecell{ \textbf{EER} \\ \textbf{Whisper} } & \makecell{ \textbf{EER} \\ \textbf{UAD} }  &  \makecell{ \textbf{EER} \\ \textbf{Whisper} } & \makecell{ \textbf{EER} \\ \textbf{UAD} } &  \makecell{ \textbf{EER} \\ \textbf{Whisper} } & \makecell{ \textbf{EER} \\ \textbf{UAD} }\\
  \hline
  Uni 1               & $t$                  & 80k  & 13.15\% & 13.15\%  & 12.90\% &  12.90\%  & 12.87\% & 12.87\% \\
  Uni 2               & $a$                  & 80k    & 10.76\% & 8.93\% & 10.78\% & 8.99\%    & 11.40\% & 8.82\% \\
  Uni 3               & $b$                  & 80k  & 33.66\% & 33.66\%  & 35.04\% & 35.04\%   & 35.70\% & 35.70\%\\
  \hline
  Multi 1               & $t$, $b$         & 80k   & 13.00\% & 13.00\%   & 12.86\% & 12.96\% & 13.08\% & 13.26\%  \\
  Multi 2               & $a$, $b$         & 80k   & 13.12\% & 10.00\%   & 14.80\% & 10.71\% & 13.90\% & 10.72\%   \\
  Multi 3               & $t$, $a$         & 80k   & 9.84\% & 9.70\%     & 9.89\% & 8.44\%   & 9.57\% & 8.27\% \\
  Multi 4               & $t$, $a$, $b$    & 80k    & 9.43\% & 9.76\%    & 9.45\% & 8.52\%   & 9.37\% & 8.55\% \\
  \hline
\end{tabular}
  \label{tab:test1}
\end{table}}
\end{document}